# Observation of Orbital-Selective Dual Modulations in an Anisotropic Antiferromagnetic Kagome Metal $TbTi_3Bi_4$

Renjie Zhang[1,2,3*], Bocheng Yu[4*], Hengxin Tan[5*], Yiwei Cheng[6,1*], Feiran Shen[7], Junye Yang[7], Dan Mu[8], Xinru Han[1,9], Alfred Zong[10] Quanxin Hu[1], Xuezhi Chen[6,1], Yudong Hu[1], Chengnuo Meng[6], Junchao Ren[6], Junqin Li[6], Zhenhua Chen[6], Zhengtai Liu[6], Mao Ye[6], Makoto Hashimoto[11], Donghui Lu[11], Shifeng Jin[2,3], Binghai Yan[12], Lunhua He[2,7,13], Ziqiang Wang[14], Tian Shang[4†], Yaobo Huang[6‡], Baiqing Lv[1,5,15§], Hong Ding[1,16,17]

*[1]Tsung-Dao Lee Institute, Shanghai Jiao Tong University, Shanghai 200240, China*
*[2] Beijing National Laboratory for Condensed Matter Physics and Institute of Physics, Chinese Academy of Sciences, Beijing, 100190, China*
*[3] University of Chinese Academy of Sciences, Beijing 100049, China*
*[4] Key Laboratory of Polar Materials and Devices (MOE), School of Physics and Electronic Science, East China Normal University, Shanghai 200241, China*
*[5] School of Physics and Astronomy, Shanghai Jiao Tong University, Shanghai 200240, China*
*[6] Shanghai Synchrotron Radiation Facility, Shanghai Advanced Research Institute, Chinese Academy of Sciences, Shanghai 201204, China*
*[7]Spallation Neutron Source Science Center, Dongguan, 523803, China*
*[8]Institute for Quantum Science and Technology, Shanghai University, Shanghai 200444, China*
*[9]Department of Physics, Southern University of Science and Technology, Shenzhen 518055, China*
*[10] Departments of Physics and of Applied Physics, Stanford University, Stanford, CA 94305, USA*
*[11] Stanford Synchrotron Radiation Lightsource, SLAC National Accelerator Laboratory, Menlo Park, California 94025, USA*
*[12] Department of Condensed Matter Physics, Weizmann Institute of Science, Rehovot 7610001, Israel*
*[13]Songshan Lake Materials Laboratory, Dongguan, 523808, China*
*[14] Department of Physics, Boston College, Chestnut Hill, Massachusetts 02467, USA*
*[15] Zhangjiang Institute for Advanced Study, Shanghai Jiao Tong University, Shanghai 200240, China*
*[16] Hefei National Laboratory, Hefei 230088, China*
*[17] New Cornerstone Science Laboratory, Shanghai 201210, China*

* These authors contributed equally to this work
† tshang@phy.ecnu.edu.cn
‡ huangyaobo@sari.ac.cn
§ baiqing@sjtu.edu.cn

## ABSTRACT

**Orbital selectivity is pivotal in dictating the phase diagrams of multiorbital systems, with prominent examples including the orbital-selective Mott phase and superconductivity. The intercalation of anisotropic layers represents an effective method for enhancing orbital selectivity and, thereby shaping the low-energy physics of multiorbital systems. Despite its potential, related experimental studies, especially those elucidating the correlation between orbital selectivity and magnetism, remain limited. In this work, we systematically examine the interplay between orbital selectivity and magnetism in the newly discovered anisotropic kagome**

**$TbTi_3Bi_4$ single crystal, and report the coexistence of orbital-selective dual band modulations ($q_1$ ~1/3 $a^*$, $q_2$ ~ 0.28 $b^*$) within the antiferromagnetic (AFM) state. By combining soft X-ray and vacuum ultraviolet angle-resolved photoemission spectroscopy measurements, neutron powder diffraction, scanning tunneling microscopy, and density functional theory calculations, we identify these dual band reconstructions as manifestations of the AFM order, driven by a (~1/3, 0.28, 0) nesting instability of the intercalated Tb 5$d_{xz}$ orbitals. These orbital-selective modulations induce unusual momentum-dependent band folding and lead to the emergence of Dirac cones only at the $\overline{\mathrm{M}}_1$ point, signaling a topological phase transition in the AFM state. Importantly, the discovery of orbital-selective (approximately 1/3, 0.28, 0) AFM order offers crucial insights into the mechanism underlying the fractional magnetization plateau in this kagome AFM metal. Our findings not only underscore the essential role of both conducting and localized electrons in determining the magnetic orders of *Ln*$Ti_3Bi_4$ (*Ln* = lanthanide) kagome metals but also offer a pathway for manipulating magnetism through selective control of anisotropic electronic structures.**

## INTRODUCTION

In multiorbital systems, the interplay among different orbitals gives rise to a diverse range of quantum phases and emergent phenomena [1–4]. Orbital-selectivity, a key feature of these systems, links specific orbitals to certain orders and quantum phases [3–5], including charge density wave (CDW), nematicity, and superconductivity. A crucial aspect of orbital selectivity is its connection with intrinsic anisotropy. For example, in transition metal dichalcogenides and tetrachalcogenides [6–9], *d*-orbitals undergo orbital-selective reconstruction within the CDW states, where the resulting orbital textures are closely related to the anisotropy. Similarly, in cuprates, the highly anisotropic stripe phase is associated with distinct behaviors of the 2$p_x$ and 2$p_y$ orbitals of oxygen [10,11]. For nematicity in the iron-based superconductors, orbital-selective occupancy of electronic states is proposed to account for the orbital orders, resulting in electronic and structural anisotropies [4,12]. In the context of superconductivity, the orbital-selective pairings have been proposed to account for the anisotropic superconducting gap across different orbitals [13,14].

The above examples not only highlight the intricate role of orbital-selectivity in determining the anisotropic correlated quantum phase, but also suggest an effective knob for manipulating the orbital characters through engineering anisotropic structures. Intuitively, introducing anisotropic structures, such as by intercalating layers, could modulate orbital distribution within the electronic structure, potentially enhancing orbital selectivity. More specifically, the anisotropic, low-dimensional Fermi surface that arises from selective orbitals might foster new correlated phases through mechanisms like Fermi surface nesting. As a result, it is anticipated that various correlated phases, including CDW and superconductivity, could emerge in anisotropic systems where orbital selectivity is significant [8,9,14]. However, the exploration of orbital selectivity and its correlation with other correlated quantum phases—especially magnetism—through the engineering of anisotropic structures, remains a relatively uncharted territory.

To understand the intricate relationship between orbital selectivity, anisotropy, and magnetic orders, we focus on a recently discovered kagome magnet, $TbTi_3Bi_4$, featuring double adjacent kagome layers interwoven with 1D zigzag chains of Tb atoms [15–17]. The unique combination of magnetic rare earth ions (4*f* and 5*d* orbitals) and kagome layers (Ti 3*d* and Bi 6*p* orbitals), positions this family of compounds as a promising platform for exploring the interplay between orbital selectivity and anisotropy. Our systematic angle-resolved photoemission spectroscopy (ARPES) measurements reveal the coexistence of dual orbital-selective band modulations in the antiferromagnetic (AFM) state: a predominant band reconstruction along zigzag chains ($q_1$ ~1/3 $a^*$) and a weaker band reconstruction perpendicular to the chains ($q_2$ ~0.28 $b^*$). Through a comprehensive analysis of the folded bands and their temperature evolution — corroborated by neutron powder diffraction (NPD), scanning tunneling microscopy (STM), and density functional theory (DFT) calculations, we attribute these modulations to the Fermi-surface-nesting-driven AFM order. Crucially, the formation of AFM order gives rise to Dirac cones at the $\overline{\mathrm{M}}_1$ point, pointing to a magnetic-order-induced topological phase transition. Following these observations, we propose a theoretical framework linking the 1/3 $a^*$ AFM phase to the unique 1/3 magnetization plateau observed in this material. Our findings not only lay the groundwork for further exploration of orbital selectivity and magnetism in similar kagome magnets but also pave the way for tuning magnetism through the control of orbital selectivity and electronic structure.

## RESULTS

$TbTi_3Bi_4$ crystallizes in space group F*mmm* (no. 69) with lattice parameters *a* = 5.8332 Å, *b* = 10.2792 Å, and *c* = 24.5883 Å [Figs. 1(a) and (b)]. It shares the same structural motif as other members in the *Ln*$Ti_3Bi_4$ family, characterized by a slightly distorted Ti double kagome layer sandwiched between Tb-Bi double layers. Notably, the Tb atoms in these Tb-Bi layers form highly anisotropic quasi-1D zigzag chains that run parallel to the *a*-axis. In addition, $TbTi_3Bi_4$ exhibits an exotic 1/3 fractional magnetization plateau when applying the magnetic field along the zigzag chain, and showcases the topological-Hall-like features [17], suggesting the intricate interplay between the electronic structure and the magnetism.

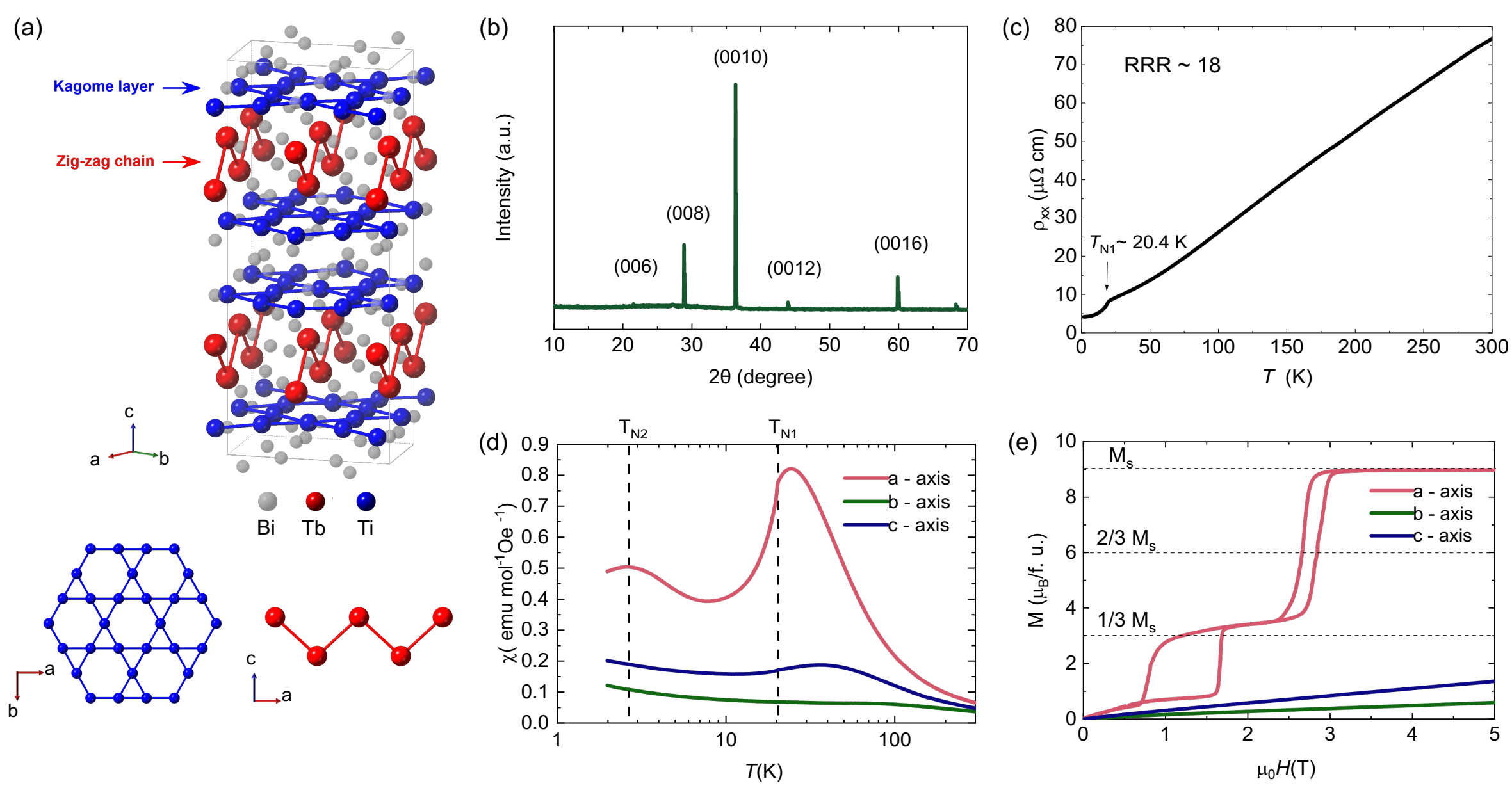

Fig. 1. Crystal structure and magnetism of $TbTi_3Bi_4$. (a) Crystal structure of $TbTi_3Bi_4$. The lower left and lower right panels show the kagome network and the intercalating zigzag chain of Tb, respectively. (b) The single crystal X-ray diffraction of $TbTi_3Bi_4$. (c) The longitudinal electrical resistance along *a*-axis, showing the AFM transition at $T_{N1}$ ~ 20.4 K. (d), (e) The temperature-dependent magnetic susceptibility (d) and the field-dependent magnetization (e) along *a*-, *b*- and *c*-axes of the $TbTi_3Bi_4$. $T_{N1}$ and $T_{N2}$ in (d) represent the corresponding magnetic transition temperatures. $M_S$, 2/3 $M_S$ and 1/3 $M_S$ in (e) indicates different magnetization states, respectively. The weak kink at 2/3 $M_S$, relating to the 2/3 magnetization plateau.

To investigate such an interplay, we systematically performed magnetic and electrical characterizations, as well as ARPES measurements. As shown in Fig. 1(c), $TbTi_3Bi_4$ exhibits metallic behavior at low temperatures, with a distinct kink in the temperature-dependent electrical resistivity $\rho_{xx}(T)$ at $T_{N1}$ ~ 20.4 K. Figure 1(d) displays the temperature dependence of magnetization $M(T)$ collected by applying magnetic fields along three different axes. For $H//a$, a sharp peak was observed at $T_{N1}$ = 20.4 K in the $M(T)$, consistent with the resistivity data in Fig. 1(c), demonstrating the onset of AFM order in $TbTi_3Bi_4$ single crystal. Furthermore, a hump-like feature emerges at $T_{N2}$ ~ 2.5 K in the susceptibility curve [Fig. 1(d)], indicative of a metamagnetic transition, consistent with other works [16,17].

Interestingly, the AFM transition is much less pronounced for either $H//b$ or $H//c$, implying a strong magnetic anisotropy in $TbTi_3Bi_4$ and suggesting that the Tb moments are predominantly aligned along the *a*-axis, i.e., the direction of the zigzag chains, with much smaller components along other directions. Furthermore, the field-dependent magnetization $M(H)$ curves [Fig. 1(e)] reinforce the magnetic anisotropy, identifying the *a*-axis as the easy axis while the *b*- and *c*-axes as the hard axes.

Uniquely, after undergoing a metamagnetic transition at approximately 1 T, there is a 1/3 magnetization plateau in the field range of approximately 1 T to 2.5 T in the $M(H)$ for $H//a$, as shown in Fig. 1(e). When the magnetic field exceeds approximately 3 T, the magnetization saturates, indicating the establishment of a forced ferromagnetic arrangement. Such metamagnetic transitions and magnetization plateau are absent for both $H//b$ and $H//c$. Magnetoresistance measurements (see Appendix A) further confirm the metamagnetic transitions observed in the $M(H)$ curves, and together with the $M(T)$ curves, they outline a rich magnetic phase diagram for $TbTi_3Bi_4$ (see Appendix A).

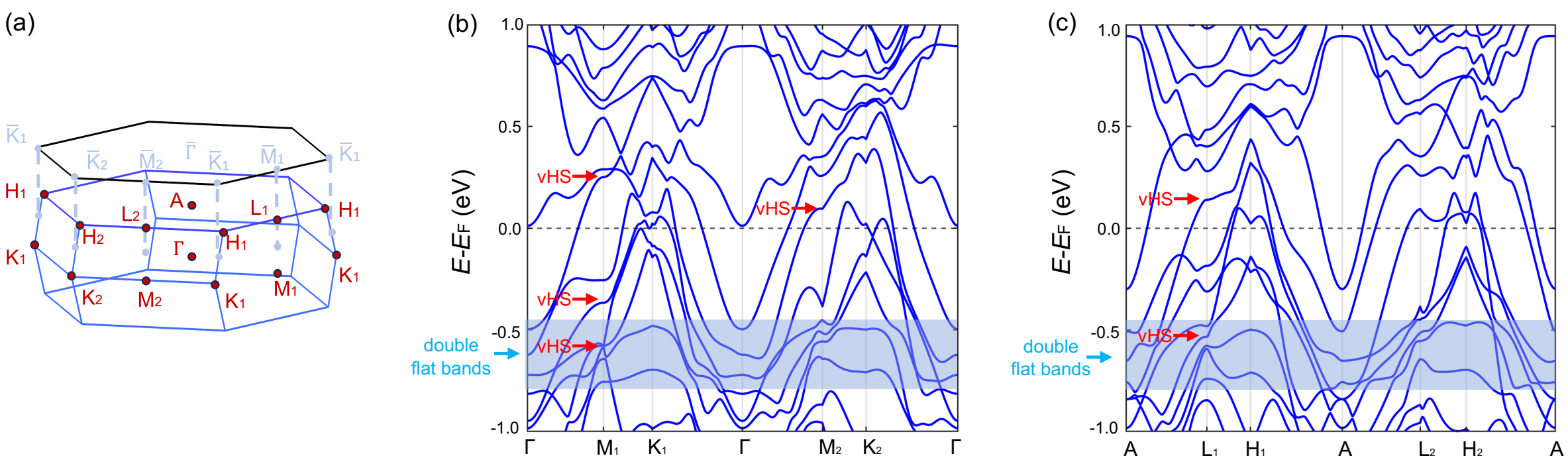

Fig. 2. The DFT calculation of the normal state. (a) Three-dimensional bulk BZ of $TbTi_3Bi_4$ (blue) and (001) projected surface BZ (black), with high symmetry points indicated. (b), (c) the DFT calculated

band structure of non-magnetic bulk state along high symmetry line in $k_z = 0$ (b) and $k_z = \pi$ (c) plane, with kagome features (vHSs, flat bands) indicated.

The observed highly anisotropic and intriguing magnetic properties are most likely due to the interplay between itinerant electrons and local magnetic moments of Tb 4*f* electrons. To elucidate the electronic structure of $TbTi_3Bi_4$, we performed DFT calculations. Fig. 2 displays the calculated band dispersion at the $k_z = 0$ and $k_z = \pi$ planes (defined in the first Brillouin zone, or BZ, same as below). The calculations reveal several characteristic features for this kagome-lattice compound, including van Hove singularities (vHS) and flat bands. Notably, the inequivalence of the vHSs at $M_1$/$M_2$ and $L_1$ highlights the anisotropy in the normal-state electronic structure, a trait commonly observed in the $LnTi_3Bi_4$ family [18,19]. Additionally, the observation of two closely spaced flat bands below the Fermi level ($E_F$) reflects the existence of double kagome layers within a single unit cell were identified [19]. Because of the presence of the rotation-symmetry-breaking intercalating layers, intrinsic strain, from the anisotropic structure, makes the Dirac point away from the high-symmetry point of the bulk BZ ($K_1$/$K_2$ and $H_1$/$H_2$) [20] and thus is absent in the DFT calculation results shown along the high symmetry line.

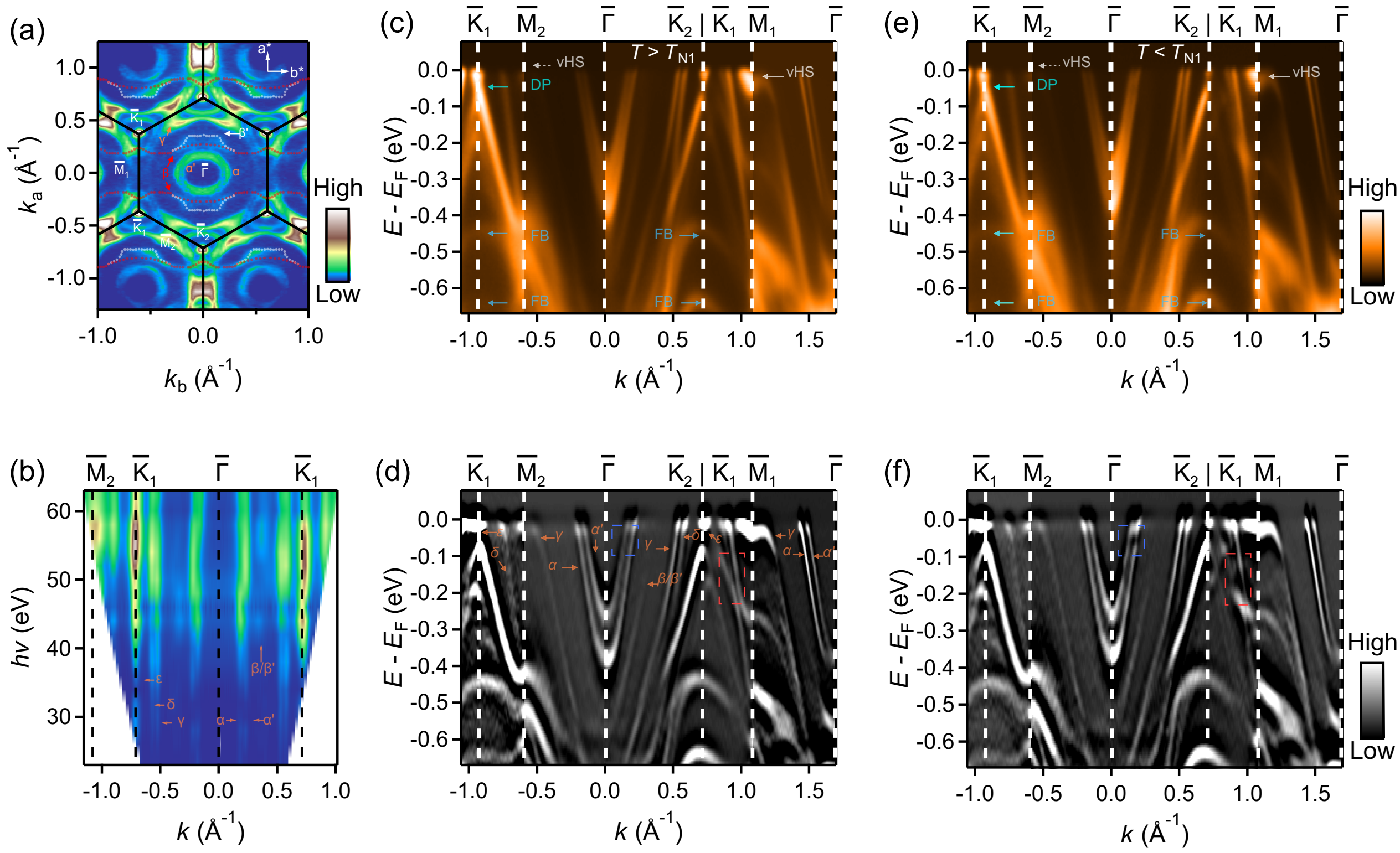

Fig. 3. The overall electronic structure of $TbTi_3Bi_4$ in the normal state and AFM state. (a) Fermi surface map of $TbTi_3Bi_4$ (using VUV light of LH, at 100 eV). The red and white dotted lines are guides for the eye indicating the quasi-1D bands. (b) Photon-energy-dependent constant energy map along $\overline{M}_2 - \overline{K}_1 - \overline{\Gamma} - \overline{K}_1$. (c), (d) Intensity (c) and the corresponding curvature plots (d) of the band structure along the high symmetry lines in the normal state ($T > T_{N1}$). “DP” means Dirac point, and “FB” means flat band. (e), (f) same as the (c) and (d), but taken in the AFM state ($T_{N2} < T < T_{N1}$). The blue dash rectangle and the red dashed rectangle in (d) and (f) indicate the band reconstruction.

To experimentally determine the electronic structure and understand its relationship with an intrinsic magnetic order, we continued to conduct ARPES measurements. We first mapped the Fermi surface within the $k_x$ - $k_y$ plane [Fig. 3(a)] and along the $k_z$ direction [Fig. 3(b)] using both vacuum ultraviolet (VUV) light and soft X-ray (Appendix C). The consistency between the VUV and soft X-ray data demonstrates that our data predominantly represent bulk states. More specifically, from Fig. 3(b), one can see that the Fermi surface along the $k_z$ direction appears nearly non-dispersive. We interpret this as a result of both the quasi-2D nature of bands near the $E_F$ and the $k_z$ broadening [21].

Next, we focused on the ARPES measurements with VUV light to clarify the band structure near $E_F$. Starting from the center of the BZ, we identify two electron-like pockets forming two concentric circular Fermi surfaces (labeled as $\alpha$ and $\alpha'$) around the $\bar{\Gamma}$ point [Fig. 3(a), (c) and (d)]. Note that the observed small splitting between $\alpha$ and $\alpha'$ is not seen in the DFT calculations (Fig. 2), and we attribute this splitting between $\alpha$ and $\alpha'$ to the surface effect caused by the self-doping of the non-neutral cleavage surface, which was commonly observed and well-studied in $AV_3Sb_5$ [22]. Next, we turn to the BZ corners, $\bar{K}_1 / \bar{K}_2$. At these points, there are point-like Fermi pockets $\varepsilon$, which is formed by the near $E_F$ Dirac point and can be clearly resolved in Fig. 3(c), (d). In addition, a triangular hole-like Fermi surface $\delta$ is seen around $\bar{K}_1 / \bar{K}_2$ [Figs. 3(a), 3(c) and 3(d)].

Interestingly, as we move toward the $\bar{M}_1 / \bar{M}_2$ points, vHSs near the $E_F$ become distinctly observable. Akin to the $AV_3Sb_5$ ($A$ = alkali metal) family [23–28], bands emanating from the vHSs cross the $E_F$ and form a large $\gamma$ pocket near the BZ boundary. However, the vHSs at $\bar{M}_1 / \bar{M}_2$ are not equivalent due to the anisotropy of this material, leading to the warped Fermi surface [18,19,29,30], which deviates from the ideal nesting condition in the $AV_3Sb_5$ family [23–28]. This imperfect nesting between vHSs may explain the absence of CDW order in the kagome layer [25].

After the identification of the typical kagome features, our attention shifts to the characteristics of the Tb 1D zigzag chains. As shown in Fig.3(a), from the Fermi surface map in the second BZ, one can identify a quasi-1D line perpendicular to the Tb chain direction (i.e., $a^*$, parallel to the $\bar{\Gamma} - \bar{M}_1$ direction). Detailed analysis (see Appendix E) reveals that this 1D line is formed by two quasi-1D Fermi surfaces $\beta$ and $\beta'$, denoted by red and white markers in Fig. 3(a), respectively. The $\beta / \beta'$ bands have an electron-like dispersion parallel to the chain direction [Fig. 3 (c) and (d)]. The presence of these quasi-1D $\beta / \beta'$ pockets, supported also by the DFT calculations (see Appendix B for details), are key features of the 1D Tb chains.

Having established the electronic structure in the normal state, we now shift our focus to the AFM state. Figures. 3(e) and (f) present the high-symmetry cuts in the AFM state ($T_{N2} < T < T_{N1}$). Compared with the normal-state results [Figs. 3(c) and (d)], we found that the kagome electronic features, including the vHSs, the Dirac points and the flat bands, remain largely unaffected by the AFM order. On the other hand, when we turn our attention to the bands along the $\bar{K}_1 - \bar{M}_1$ path [highlighted by the red dashed rectangles in Fig. 3(d) and (f), aligned with the zigzag chain direction], we observe clear differences, suggesting the emergence of momentum-dependent band reconstruction in the AFM state.

To better visualize the band reconstruction, we systematically compared the measured high-resolution intensity and the corresponding curvature intensity plots along both $\bar{K}_1 - \bar{M}_1 - \bar{K}_1$ [Figs. 4(a) and (b)] and $\bar{K}_2 - \bar{M}_2 - \bar{K}_1$ directions [Figs. 4(c) and (d)]. Our findings are twofold. Firstly, a distinct Dirac-cone-type crossing and some additional new states were observed along $\bar{K}_1 - \bar{M}_1 - \bar{K}_1$ direction below $T_{N1}$, as demonstrated by the red markers in Fig. 4(b)(ii) and the emergence of Peak 3 in Fig. 4(g). Secondly, a suppression of the band intensity is noticeable in the energy range of approximately -0.1 eV to -0.2 eV along the $\overline{\mathrm{M}}_1-\overline{\mathrm{K}}_1$ direction below $T_{\mathrm{N1}}$, more clearly in the curvature intensity plot [highlighted by the black rectangle in Fig. 4(b)(i)]. In sharp contrast, despite these pronounced changes around $\overline{\mathrm{M}}_1$, there is no distinct variations at $\overline{\mathrm{M}}_2$ along the $\overline{\mathrm{K}}_2-\overline{\mathrm{M}}_2-\overline{\mathrm{K}}_1$

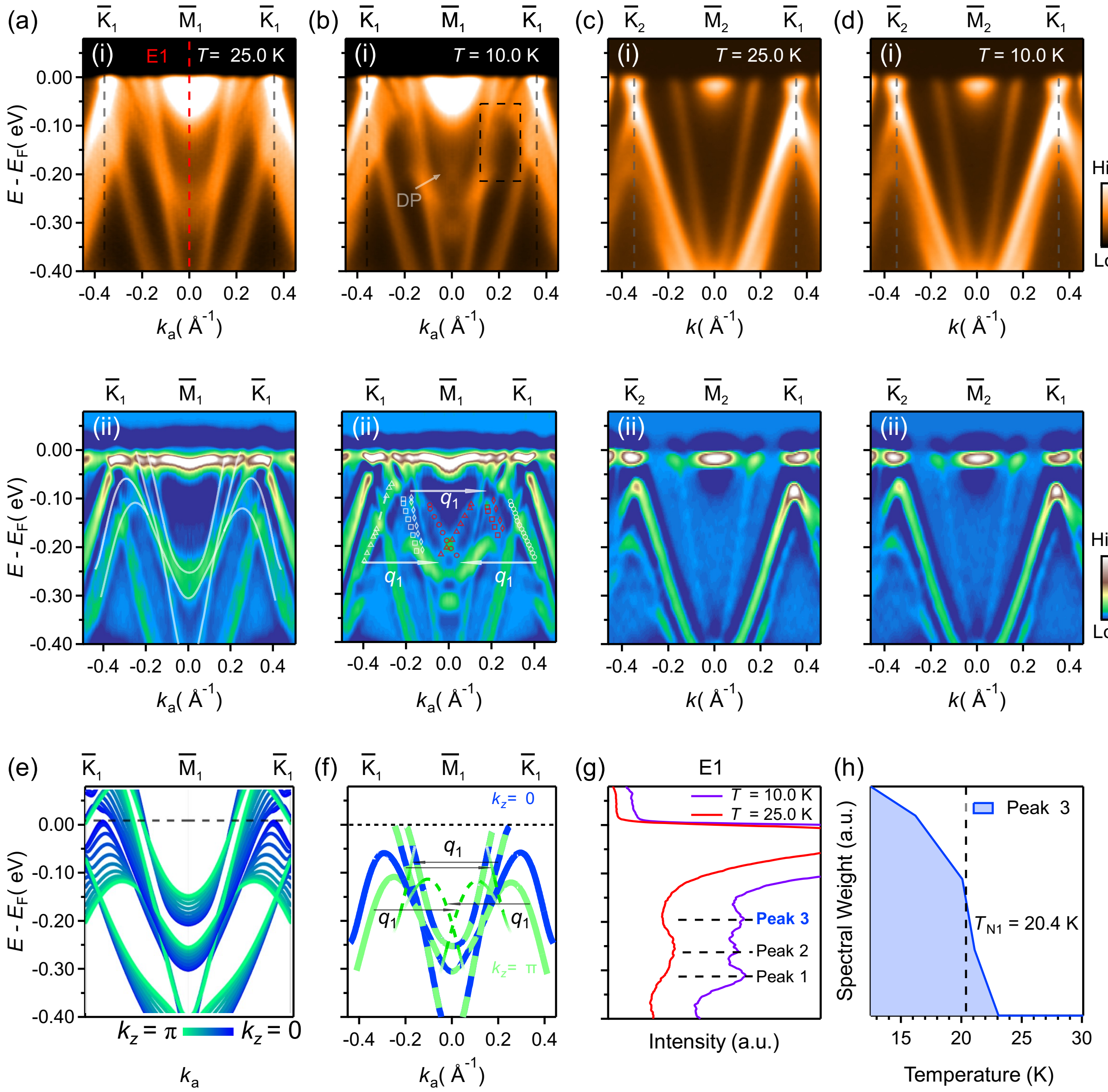

direction [Figs. 4(c) and (d)] and $\overline{\mathrm{M}}_2-\overline{\Gamma}-\overline{\mathrm{M}}_2$ (Appendix D, Fig. 14), underscoring the strong anisotropy and momentum dependence for the band reconstruction.

Fig. 4. Temperature dependence of the electronic band structure at $\overline{\mathrm{M}}_1$ (along *a* direction) and $\overline{\mathrm{M}}_2$. (a) Intensity (i) and the corresponding curvature (ii) plots along $\overline{\mathrm{K}}_1-\overline{\mathrm{M}}_1-\overline{\mathrm{K}}_1$ direction in the normal state ($T$ = 25.0 K). E1 indicates the EDC at $\overline{\mathrm{M}}_1$, and the comparison with the EDC at $\overline{\mathrm{M}}_1$ in AFM state is shown in Fig. 4(g). (b) Same as (a), but taken in the AFM state ($T$ = 10.0 K). The black dashed rectangle in (i) highlights the DOS suppression. "DP" in (i) represents the Dirac point. Red marks in (ii) indicate the folding bands, and white marks in (ii) are the respective original bands. The white arrow indicates the folding vector $q_1$. (c), (d) Same as (a) and (b), but taken along $\overline{\mathrm{K}}_2-\overline{\mathrm{M}}_2-\overline{\mathrm{K}}_1$

direction. (e) DFT calculation of $k_z$-projected band structure along $\overline{\mathrm{K}}_1 - \overline{\mathrm{M}}_1 - \overline{\mathrm{K}}_1$ in the normal state. (f) Band dispersion extracted from (a), thick green and blue lines represent $k_z = \pi$ and $k_z = 0$ bands, respectively. Green dashed lines represent folding bands. (g) Energy distribution curves (EDCs) taken from $\overline{\mathrm{M}}_1$ (indicated by E1) in the intensity plots of (a) and (b). Peak 3 represents the Dirac crossing point. (h) Temperature evolution of the spectral weight of Peak 3, data processing details are described in Appendix D.

To be more precise, we extracted the dispersions of both the original bands in the normal state [indicated by gray lines in Fig. 4(a)] and the emergent Dirac bands in the AFM state [marked by red squares and diamonds in Fig. 4(b)-(ii)]. Intriguingly, these emergent bands can be folded onto the original bands by a vector $q_1 \approx 0.36$ Å$^{-1}$ [approximately 1/3 $a^*$, where $a^* = 2\pi/a$, as shown by the black arrows in Fig. 4(f)], suggesting the emergence of a density wave order associated with the vector $q_1$. Importantly, according to the irreducible representations of the little group at $\mathrm{L}_1$ and $\mathrm{M}_1$ in space group F*mmm*, no symmetry-enforced degeneracies are expected at these points [31]. Thus, the emergent gapless Dirac point at the BZ boundary must either be protected by magnetic symmetries or originate from surface states induced by bulk band inversions. In either scenario, the presence of a Dirac-cone-like crossing points to an underlying topological state in the AFM phase.

Away from $\overline{\mathrm{M}}_1$, the formation of the density wave order gives rise to hybridization gaps at crossing points of the folded bands, accounting for the suppression of spectral intensity in the region highlighted by the black rectangle in Fig. 4(b). Furthermore, a comparison between the experimentally observed reconstructed bands and the DFT-calculated $k_z$-projected band structure reveals that these reconstructions are predominantly concentrated in the $k_z = \pi$ plane. This suggests that the $k_z = \pi$ plane plays a particularly significant role in the band reconstruction process.

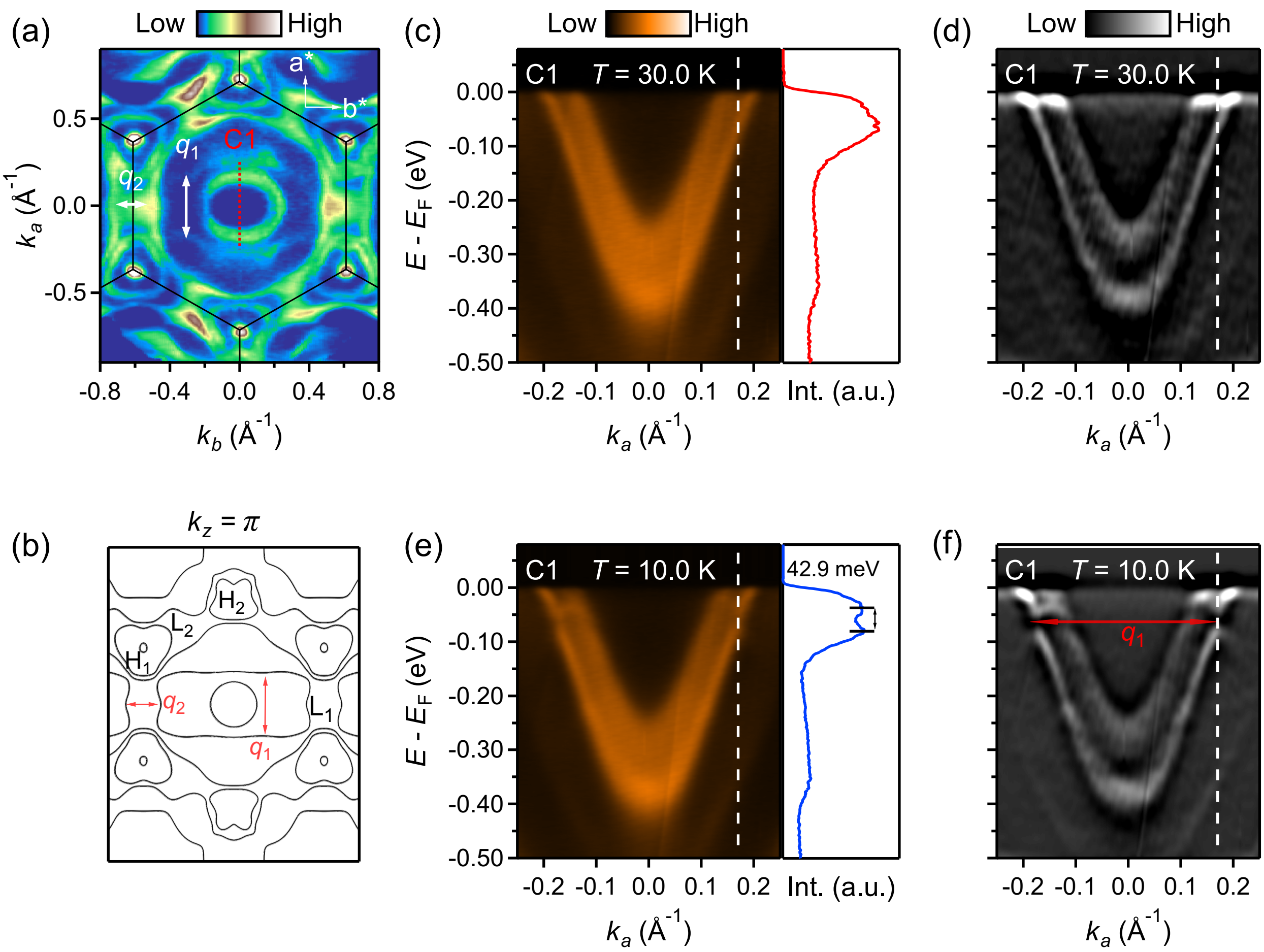

Fig. 5. Band folding around $\bar{\Gamma}$ point. (a) Fermi surface map taken with photons of LV+LH at 60 eV, with the red dashed line indicating the momentum location of cut C1 shown in (c)-(f), white arrows represent nesting vectors $q_1$ and $q_2$ connecting two pairs of parallel Fermi surfaces, respectively. (b) DFT calculated Fermi surface in the $k_z = \pi$ plane. $q_1$, $q_2$ are identical to those shown in (a). (c), (d) Intensity (c) and the corresponding curvature (d) plots of $\alpha / \alpha'$ pockets ($T = 30.0$ K) taken along C1 [red line in (a)]. EDC at the white dash line is present in the right panel of (c). (e), (f) Same as (c) and (d), but taken at $T = 10.0$ K. Gap opening (~ 42.9 meV) is visualized in the EDC shown in the right panel of (e).

To further verify the band folding along $a^*$, we turn our attention to the electron pockets near the $\bar{\Gamma}$ point. The measured intensity plots above and below $T_{N1}$ along the $\bar{\Gamma}-\bar{K}_2$ directions are summarized in Fig. 5. A gap of approximately 42.9 meV was observed on the outer band around $\bar{\Gamma}$, with obvious band back-bending features [Fig. 5 (e) and (f)]. As expected, the momentum separation of the left and right sides of the gapped outer band matches well with $q_1$. These observations reaffirm the existence of density wave modulations in $TbTi_3Bi_4$.

One of the predominating driving forces behind the density wave modulation is the Fermi surface nesting. In quasi-1D systems, the ideal nesting condition is often naturally present due to the prevalence of 1D Fermi surface. As demonstrated in Fig. 3(a) and Appendix E, $TbTi_3Bi_4$ hosts a quasi-1D Fermi surface $\beta$ along $\bar{\Gamma}\bar{M}_1$ thanks to the intercalation of 1D Tb chains. Therefore, it is natural to link the density wave to the $\beta$ Fermi surface. Indeed, from the mapped Fermi surfaces [Fig. 5(a)], one can see that the two branches of the $\beta$ Fermi surface can be connected by a vector $q_1$ along $a^*$, strongly suggesting $\beta$ as the underlying driving force for the observed density waves. Meanwhile, one would expect the opening of an energy gap at the $E_F$ for this $\beta$ pocket. To this end, we performed a cut along $\beta$ (see Appendix E). Comparing the AFM state with the normal state, we indeed find a gap opening of approximately 77.6 meV with respect to the $E_F$. The above observation is further supported by our DFT calculations. As shown in Fig. 5(b), a large, well-nested portion of quasi-one-dimensional Fermi surfaces is found under the wavevector $q_1$ in the $k_z = \pi$ plane, consistent with our ARPES results. These observations pin down that the anisotropic band reconstruction observed below $T_{N1}$ is primarily driven by the nesting of quasi-1D Fermi surface.

After confirming the band reconstruction at specific points ($\overline{\Gamma}$ and $\overline{M}_1$) in the $a^*$ direction, a natural question arises regarding the universality of Fermi surface nesting: Can it also occur in other directions? Notably, another pair of parallel Fermi surfaces exists around $L_1$ (projected onto $\overline{M}_1$ in a surface BZ) along $a^*$ connected by wavevector $q_2$ [Figs. 5 (a) and 5(b)]. To investigate the Fermi surface nesting in the $b^*$ direction, we conducted a temperature-dependent ARPES measurement along $\overline{\Gamma}$-$\overline{M}_1$-$\overline{\Gamma}$ (Fig. 6). A comparison between Fig. 6(a) ($T > T_{N1}$) and (b) ($T < T_{N1}$) clearly reveals modifications in the energy bands around $\overline{M}_1$. Specifically, a notable change is the band folding associated with $q_2 = 0.28\ b^*$ ($b^* = 2\pi/b$) below $T_{N1}$ [Figs. 6(b)(i) and 6(b)(ii)]. Interestingly, $q_2$ directly connects the nested Fermi surfaces near $\overline{M}_1$. Collaboratively, Fig. 6(b)(ii) exhibits a back-bending feature in the near-$E_F$ band at $\overline{M}_1$, in contrast to the normal-state band structure shown in Fig. 6(a)(ii). To better visualize the gap opening, we divided the spectra by a Fermi-Dirac function convolved with the experimental energy resolution [Figs. 6(a)(iii) and 6(b)(iii)]. A gap is clearly resolved only on the inner pocket centered at $\overline{M}_1$, whereas other near-$E_F$ bands around $\overline{M}_1$ and $\overline{\Gamma}$ remain largely unaffected. This provides strong evidence for the existence of density wave order along $b^*$.

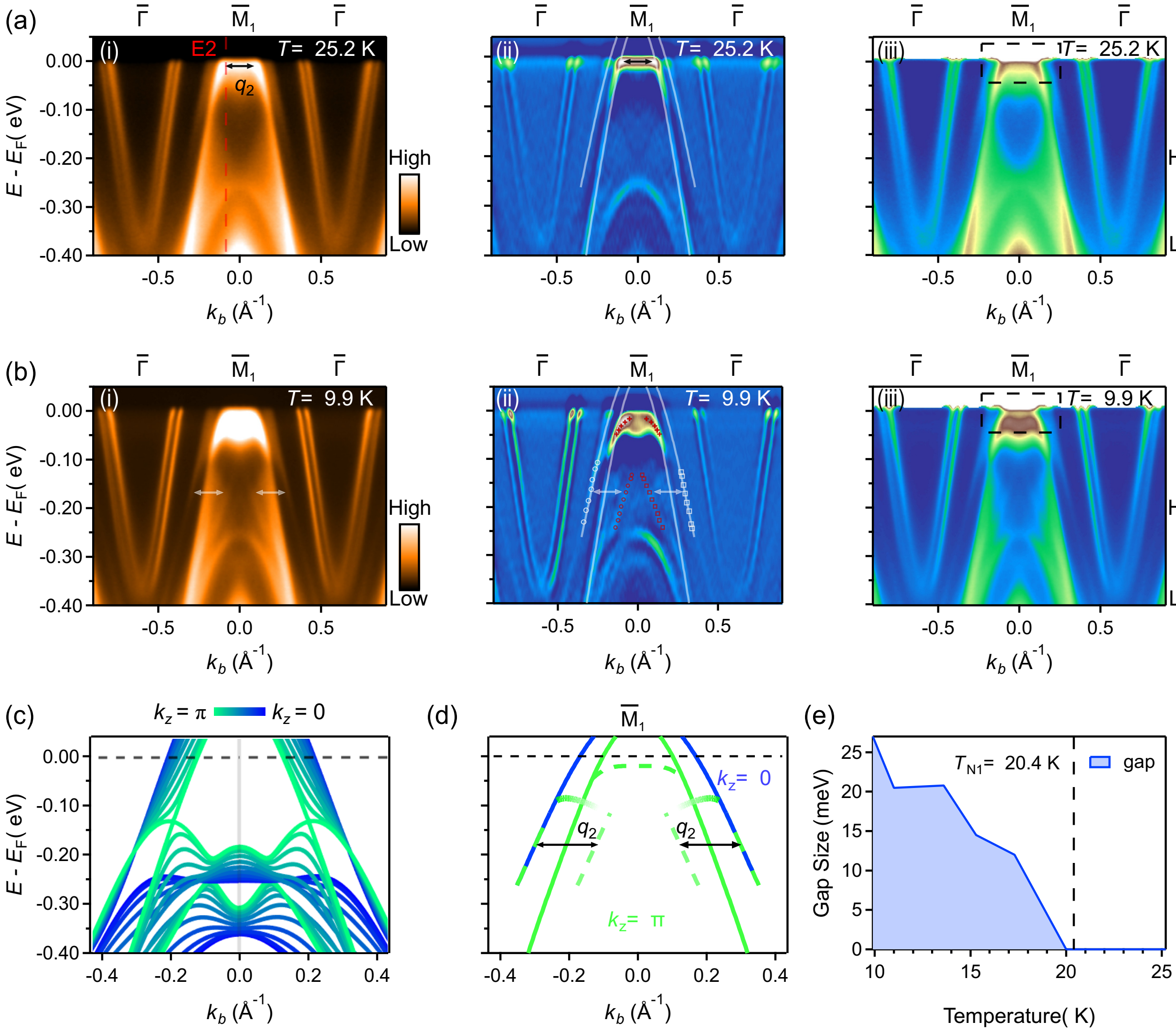

Fig. 6. Temperature dependence of the electronic band structure at $\overline{\mathrm{M}}_1$ (along $b^*$ direction). (a) ARPES intensity plot (i), curvature plot (ii) and intensity plot divided by Fermi-Dirac function convolved with energy resolution (iii), in the normal state (T = 25.2 K). The red dashed line in (i) indicates the EDC, at Fermi surface, shown in (e). The arrow represents the wavevector, $q_2$, connecting Fermi surfaces. White curves in (ii) indicate the extracted energy bands near $E_F$. (b) Same as (a), but measured in AFM state (T = 9.9 K). The white and red marks in (ii) represent the original and folding bands, respectively, connected by the folding vector $q_2$. The dashed rectangle box in (iii) highlights the gap opening due to the Fermi surface nesting. (c) $k_z$-projected band structure along $\overline{\Gamma}$-$\overline{\mathrm{M}}_1$-$\overline{\Gamma}$, obtained from DFT calculation. (d) Band structure along $\overline{\Gamma}$-$\overline{\mathrm{M}}_1$-$\overline{\Gamma}$, extracted from (a)-(ii). Dashed curves indicate the reconstructed bands. (e) Temperature-dependent energy gap taken from symmetrized EDC of E2 in (a)-(i).

To gain a comprehensive understanding of the electronic mechanism underlying band reconstruction near $\overline{\mathrm{M}}_1$ along $\overline{\Gamma}$-$\overline{\mathrm{M}}_1$-$\overline{\Gamma}$, we plotted $k_z$-projected band structure [Fig. 6(c)] and extracted the experimentally observed energy bands [Fig. 6(d)]. In line with the band reconstructions along $\overline{\mathrm{K}}_1$-$\overline{\mathrm{M}}_1$-$\overline{\mathrm{K}}_1$ (Fig. 4), the primary modifications of energy bands along $b^*$ direction also mainly occur in the $k_z$

= π plane, where a well-nested rectangular quasi-1D Fermi surface is evident [Figs. 5(a) and 5(b)]. The symmetrized energy distributed curve (EDC) at $k_F$, marked as E2 in Fig 6(a)(i) demonstrate that the gap on Fermi surface in AFM state ($T$ = 9.9 K) is approximately 27 meV [Appendix D, Fig 13(c)], lower than the corresponding gap value of 77.6 meV in the $a^*$ direction (Appendix E), highlighting the strong anisotropy of the gap on the quasi-1D Fermi surface. Furthermore, Fig 6(e) shows that the gap closing temperature is around 20.0 K, consistent with $T_{N1}$.

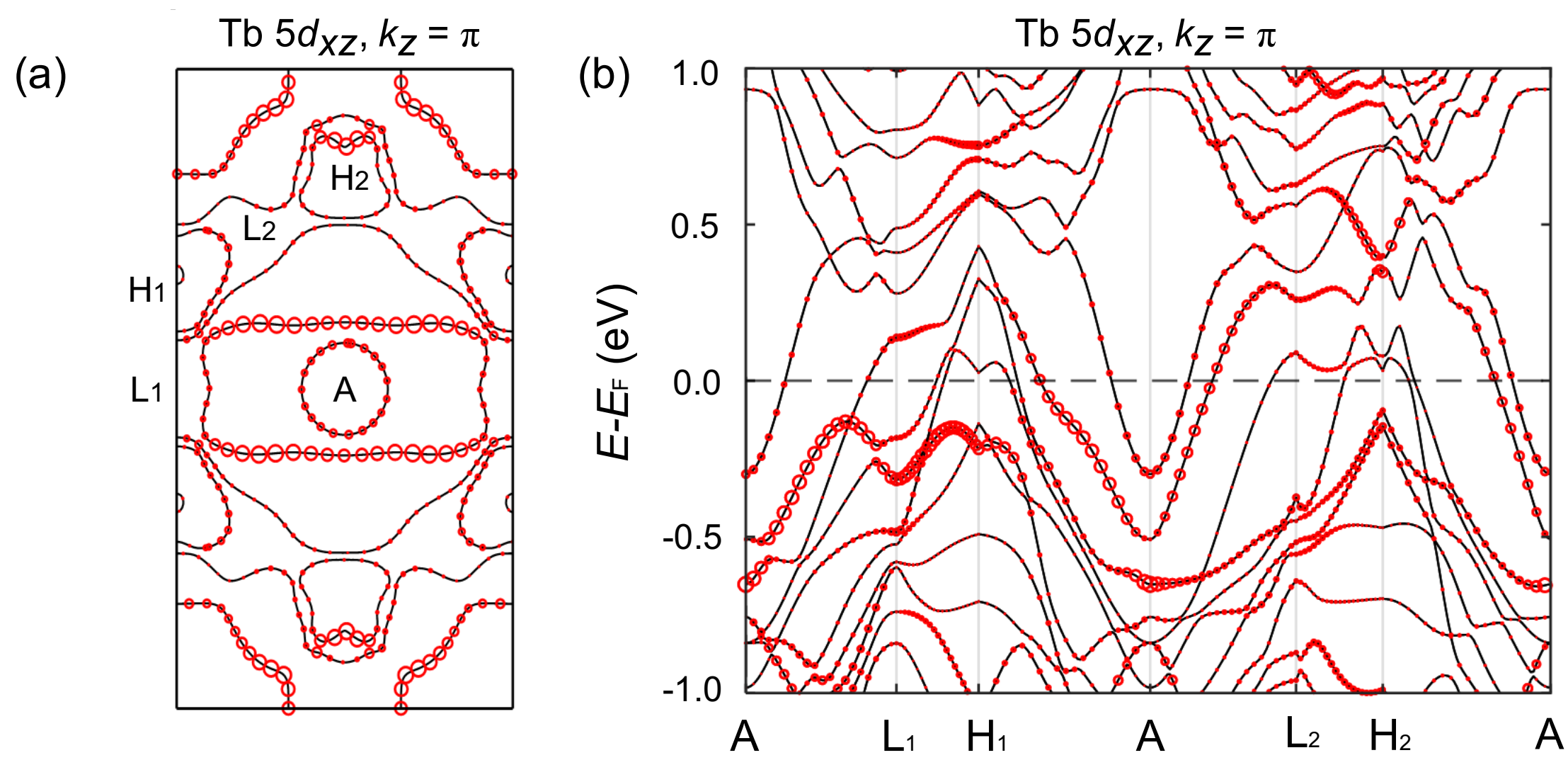

Fig. 7. DFT calculated orbital distribution of Tb-5$d_{xz}$. (a) DFT calculated orbital distribution of Tb-5$d_{xz}$ along the Fermi surface in the $k_z$ = π plane. (b) DFT calculated orbital distribution of Tb-5$d_{xz}$ in the energy band in the $k_z$ = π plane. The size of red circles indicates the weight of Tb-5$d_{xz}$, the black lines indicate the calculated Fermi surface (a) and the band structure (b), respectively.

Given the multiorbital nature of $TbTi_3Bi_4$, understanding the orbital distribution is essential for comprehending the above discovery of momentum-dependent dual band reconstructions. Since the band reconstruction is most pronounced in the $k_z$ = π plane [Figs. 4(e) and 4(f), 6(c) and 6(d)] and along the quasi-1D Fermi surface, our analysis focuses on these features. Figure 7 and Appendix F summarize the DFT results on the orbital character. Interestingly, though the Ti 3$d$ orbitals dominate the near-$E_F$ energy bands (see Appendix F), the 5$d_{xz}$ orbital ($x$ and $z$ align with $a$ and $c$ axis, respectively) of Tb is closely tied to the reconstructed bands around $L_1$ ($\overline{M}_1$) and A ($\overline{\Gamma}$) below $T_{N1}$ (Fig. 7). Notably, we observed that 5$d_{xz}$ orbital component is primarily concentrated on the portion of the quasi-1D Fermi surface parallel to the A-$L_1$ ($\overline{\Gamma}$ - $\overline{M}_1$) direction, rather than being evenly distributed across the entire Fermi surface. This naturally explains the strong anisotropy in the gap along A-$H_2$ ($\overline{\Gamma}$ - $\overline{K}_2$) and A-$L_1$ ($\overline{\Gamma}$ - $\overline{M}_1$), demonstrating the anisotropic orbital-selective dual modulation in the AFM state. To experimentally verify the orbital characters, we conducted polarization-dependent ARPES measurements on the β band along both the $\overline{\Gamma}$ - $\overline{K}_2$ and $\overline{\Gamma}$ - $\overline{M}_1$ directions, which is compatible with DFT calculation results (Appendix G).

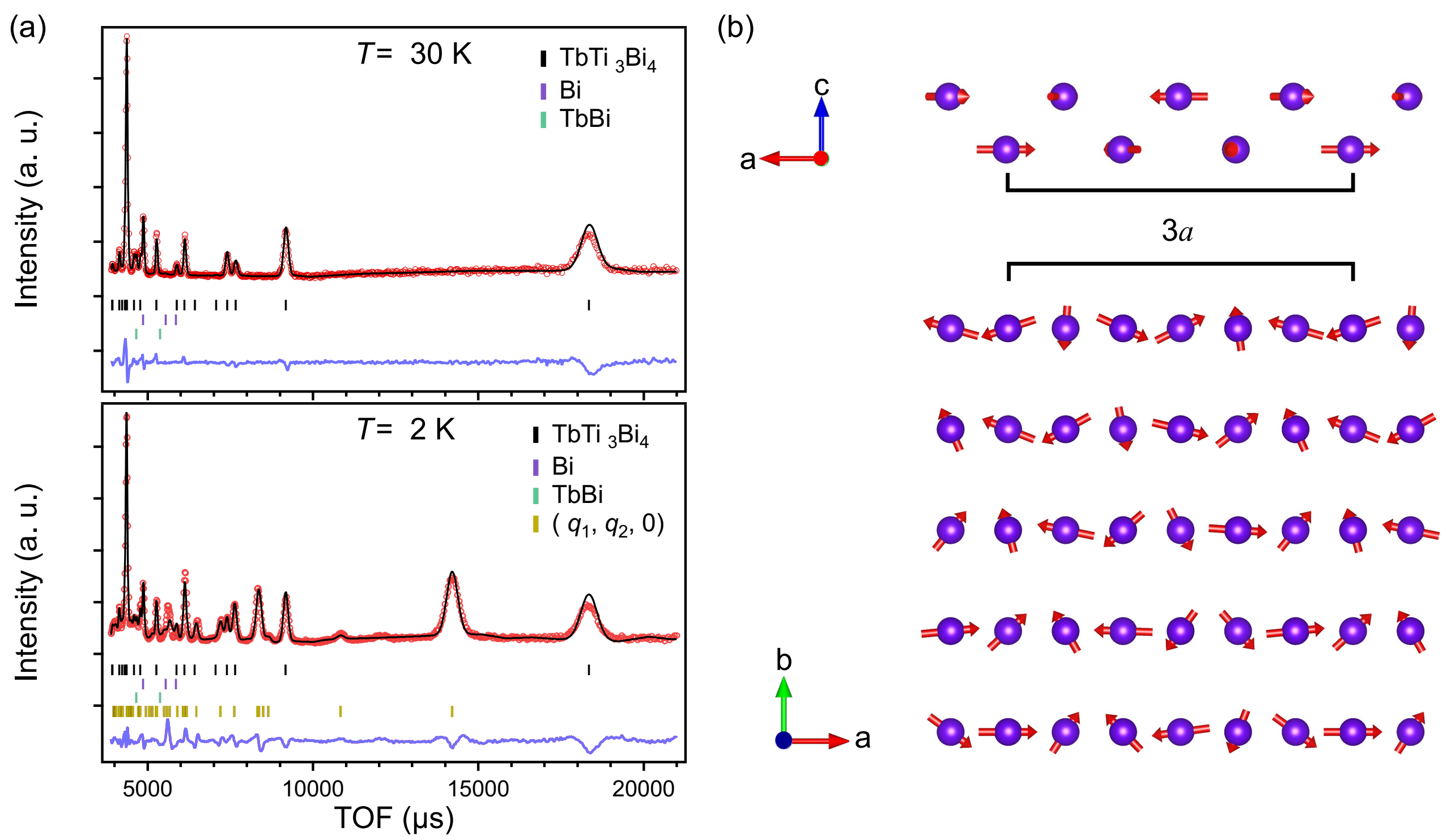

Fig. 8. Neutron powder diffraction (NPD) pattern and a possible magnetic structure corresponding to the vector ($q_1$, $q_2$, 0). (a) upper and lower panel present the NPD pattern in normal and AFM state, respectively. Black, purple, green and yellow ticks represent diffraction peak positions of $TbTi_3Bi_4$, Bi, TbBi and ($q_1$, $q_2$, 0) magnetic structure, respectively. The blue lines at the bottom indicate the residual error between fit and experimental results. (b) A proposed magnetic structure with magnetic vector ($q_1$, $q_2$, 0). Along the zigzag chain (*a*-axis), the period is approximately 3*a*, while with an incommensurate arrangement of spins along *b*-axis. Top: zigzag chain of Tb, with approximate 3*a*-period; Bottom: the Tb layer.

Given the consistency between the onset temperature of magnetic transition and band reconstruction, it is reasonable to associate the anisotropic Fermi surface nesting with the novel magnetism. To confirm this, we conducted NPD measurements both above and below $T_{N1}$ to directly characterize the magnetic structure [Fig. 8(a)]. Evidently, new features emerge in the AFM phase, suggesting the emergence of a novel magnetic structure. Through Rietveld refinement of the NPD patterns, we found that the new peaks can be well-described by a magnetic structure with a wavevector of ($q_1$, $q_2$, 0), where $q_1 = 0.35\ a^* \sim 1/3\ a^*$ and $q_2 = 0.28\ b^*$. This wavevector is consistent with the nesting vector obtained from ARPES measurements. Additionally, our STM measurements performed above and below $T_{N1}$ show no evidence of charge order with the wavevector (1/3, 0.28, 0) (see Appendix H), thereby ruling out CDW order as the predominant order. Based on the wavevector (0.35, 0.28, 0) and the Rietveld refinement, we propose a possible magnetic strucure characterized by an approximate triple-period ($\sim 1/3\ a^*$) along the *a*-axis and an incommensurate arrangement along the *b*-axis [Fig. 8(b)]. The moment magnitudes exhibit only weak, gradual modulation. We note that Fig. 8(b) presents one plausible solution derived from the NPD data; alternative magnetic configurations, e.g. spin helix, however, cannot be fully excluded.

## DISCUSSION

Taken together, our temperature-dependent ARPES results, complemented by NPD, STM, and DFT calculations, reveal orbital-selective dual modulations in the band structure characterized by a wavevector of (1/3, 0.28, 0), originated from an AFM state. The following key observations support this interpretation: (1) The onset of band reconstruction — such as the folding of Dirac bands (see peak 3 in Figs. 4(g) and 4(h)) — occurs at the bulk AFM transition temperature ($T_{N1} \sim 20.4$ K); (2) NPD, which is sensitive to magnetic order, detects long-range magnetic ordering below $T_{N1}$. The observed magnetic structure shares the same wavevector as that of the Fermi surface nesting, strongly indicating an AFM state driven by electronic instability; (3) The reconstructed quasi-1D Fermi surface predominantly involves the $5d_{xz}$ orbitals of Tb atoms, which carry strong $4f$ magnetic moments and participate in the long-range AFM order; (4) STM measurements (see Appendix H) reveal no evidence of charge ordering or structural transitions with the modulation vector (1/3, 0.28, 0), thereby confirming the magnetic origin of the orbital-selective modulation. Furthermore, a structural transition would typically induce folding across all electronic bands. However, bands near the $\overline{\mathrm{M}}_2$ and $\overline{\mathrm{K}}_1 / \overline{\mathrm{K}}_2$ points remain largely unaffected (Fig. 4 and Fig. 6), further supporting a magnetic — rather than structural — origin of the band reconstruction.

We now examine the key mechanisms driving the emergent magnetism in the $Ln\mathrm{Ti_3Bi_4}$ family. First of all, the Ruderman-Kittel-Kasuya-Yosida (RKKY) interaction is a well-established mechanism for generating diverse magnetic ground states, including helical spin-density waves and skyrmion lattices [32–34]. In the $Ln\mathrm{Ti_3Bi_4}$ compounds, we demonstrate for the first time that RKKY interactions—enhanced by quasi-1D Fermi surface nesting—play a central role in determining the distinct magnetic behaviors observed across the series. As noted by Ortiz *et al.* [35], $\mathrm{EuTi_3Bi_4}$ and $\mathrm{GdTi_3Bi_4}$ contain magnetic ions with identical total angular momentum ($J = S = 7/2$), yet exhibit fundamentally different magnetic orders: $\mathrm{EuTi_3Bi_4}$ is ferromagnetic, whereas $\mathrm{GdTi_3Bi_4}$ displays antiferromagnetism with a 1/3 magnetization plateau. This contrast correlates with their underlying electronic structures—$\mathrm{GdTi_3Bi_4}$, like $\mathrm{TbTi_3Bi_4}$, possesses a well-nested quasi-1D Fermi surface, while $\mathrm{EuTi_3Bi_4}$ does not [29]. These findings demonstrate the critical role of Fermi surface geometry, particularly nesting instabilities, in mediating magnetic interactions. They further highlight low-dimensional structural motifs (e.g. zigzag chains) as promising design principles for engineering unconventional magnetic states.

In addition to Fermiology, orbital character plays an equally crucial role in shaping the magnetic behavior of the $Ln\mathrm{Ti_3Bi_4}$ series. Within the framework proposed by Campbell [36], exchange interactions involving $4f$–$5d$ and $5d$–$5d$ (or $3d$–$5d$) orbitals are central to the magnetism of rare-earth/transition-metal systems. Experimental studies on multi-orbital interfaces, such as $\mathrm{LaAlO_3/EuTiO_3/SrTiO_3}$ heterostructures, have revealed orbital-selective magnetic correlations [37], where the $3d_{xz/yz}$ orbitals mediate ferromagnetic coupling between $\mathrm{Eu^{2+}}$ ions, while the $3d_{xy}$ orbital promotes antiferromagnetic interactions. Complementary theoretical work has further emphasized the importance of orbital-dependent exchange in stabilizing skyrmion textures in Gd-based compounds [38]. In $\mathrm{TbTi_3Bi_4}$, we reveal an orbital-selective dual modulation, in which the Tb $5d_{xz}$ orbital plays a direct role in establishing the antiferromagnetic ground state. This finding highlights the essential contribution of orbital degrees of freedom and supports the view that orbital selectivity, alongside Fermi surface nesting, is a key ingredient in driving the rich magnetism of the $Ln\mathrm{Ti_3Bi_4}$ family.

Finally, it is worth noting that local 4*f* moments and their associated anisotropy, shaped by the crystalline electric field (CEF), are indispensable to the full magnetic landscape of $LnTi_3Bi_4$. In $TbTi_3Bi_4$, the pronounced magnetization anisotropy has been attributed to strong CEF splitting of the $Tb^{3+}$ ion levels. This stands in stark contrast to the nearly isotropic magnetic behavior in $GdTi_3Bi_4$, where the half-filled 4*f* shell minimizes CEF effects [15]. These observations demonstrate the necessity of incorporating CEF effects into any comprehensive understanding of magnetism in the $LnTi_3Bi_4$ series.

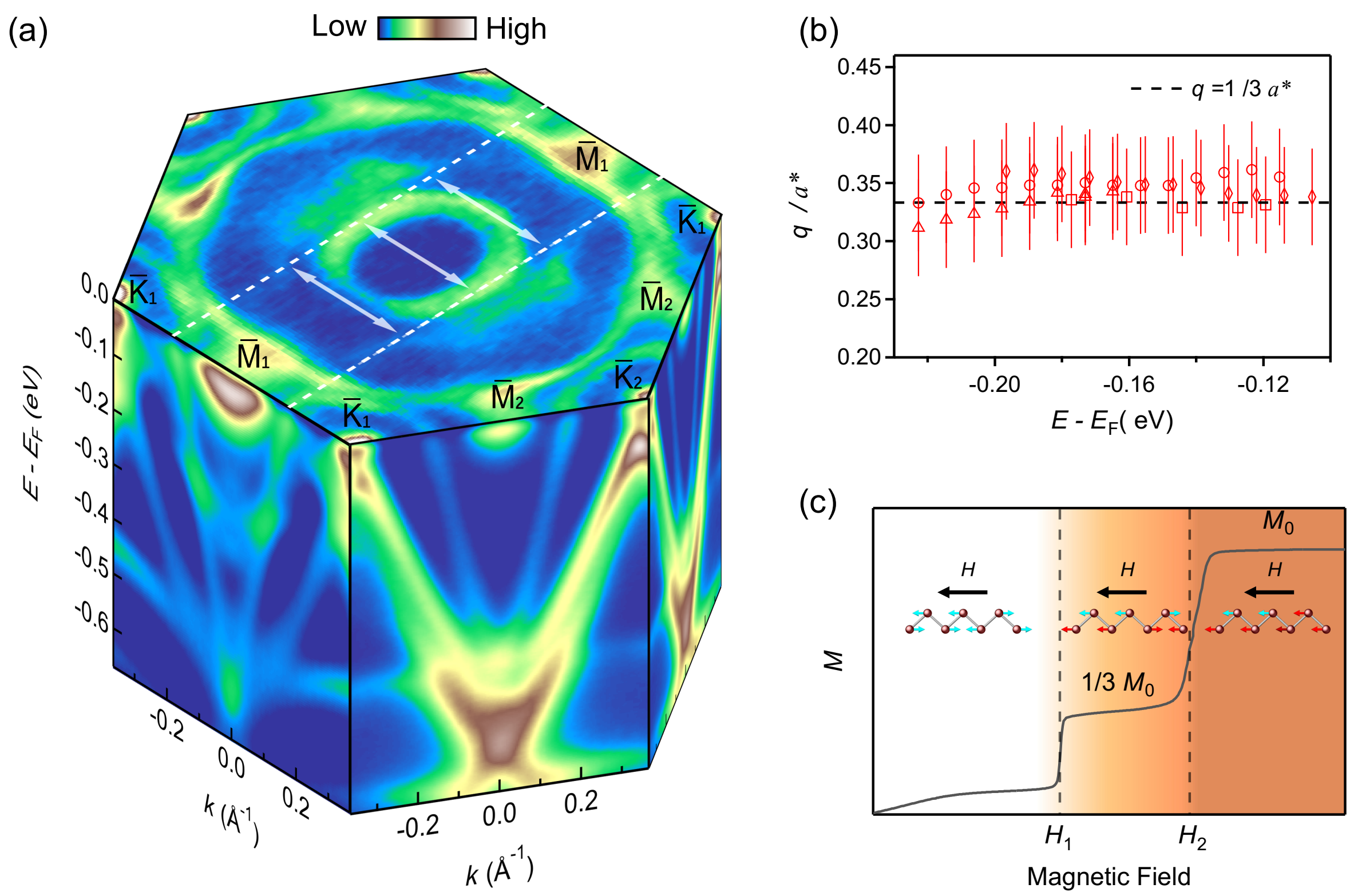

Fig. 9. Summary of electronic and magnetic properties of $TbTi_3Bi_4$. (a) Anisotropic electronic structure and reconstruction in the AFM state. White arrows indicate the nesting vector. The obvious anisotropic reconstruction is clearly visualized on the side surfaces of $\overline{M}_1$, and on the contrary, $\overline{M}_2$ does not show observable changes in the AFM state. (b) all the *q*-vectors taken from Fig. 4(d) and Fig. 5(f), these vectors are approximately around 1/3 $a^*$, the error bar is estimated from the FWHM of the Dirac point in Fig. 4(b). (c) schematic of the mechanism of 1/3 fractional magnetization plateau based on 3*a* AFM state (intralayer coupling larger than interlayer coupling). Red colored arrows indicate the flipped spin, dark red arrows represent the spin flipped twice. $H_1$ and $H_2$ indicate the magnetic field of metamagnetic transition.

More importantly, the above analysis leads us to propose a plausible mechanism for explaining the 1/3 fractional magnetization plateau based on the unique (0.35,0.28,0) AFM state. As illustrated in Fig. 8, the magnetic structure exhibits a special modulation along both the *a*- and *b*-axes, resulting in a 2D configuration. However, as mentioned earlier, this system inherently possesses a dichotomy along the

*a*- and *b*-axes in three key aspects: First, the material has a quasi-1D lattice structure, which generates an anisotropic CEF that significantly impacts the local magnetic moments; Second, as shown by the magnetization behavior, $TbTi_3Bi_4$ exhibits an easy-axis along the *a*-axis, with much weaker magnetization along the *b*- and *c*-axes. More specifically, recent single-crystal neutron diffraction measurements on $TbTi_3Bi_4$ reveal that the magnetic moments along the *a*-axis are approximately 3 times larger than those along the *b*-axis [17]; Third, in terms of the electronic structure, our ARPES measurements indicate that the energy gaps resulting from Fermi surface nesting along the $a^*$ and $b^*$ directions exhibit strong anisotropy. Similar to the magnetic moments observed by single-crystal neutron diffraction, the gap size along the $a^*$ direction is approximately 3 times that along the $b^*$ direction. Given the strong anisotropy in the lattice structure, magnetism, and electronic properties, we believe that the response of the AFM state to an external magnetic field is dramatically different along the *a*- and *b*-axes. The modulation associated with $q_2$ along the *b*-axis (interchain) is likely suppressed when a field is applied along the *a*-axis. Therefore, within a moderate field range (e.g., between 0 and $H_1$ in Fig. 9(c)), the moments along the *b*-axis will be eliminated, and all magnetic moments will align or anti-align along the *a*-axis, maintaining a periodicity of approximately $3a$ along the *a*-axis due to the relatively large gap in the $q_1$ modulation along the $a^*$ direction. Although magnetic moments may exhibit some components along the c-axis, we believe these are negligible and can be ignored. Since the $q_1$-vector is approximately 1/3 $a^*$, the possible magnetic structure should have a period of approximately $3a$ in real space. In this model [Fig. 9 (c)], each zigzag chain constitutes of two layers of local moments. In each layer, moments are arranged as left-left-right (or right-right-left), with adjacent layers oriented oppositely. Assuming that intralayer coupling is stronger than the interlayer coupling (an alternative case is illustrated in Appendix I), an applied magnetic field exceeding $H_1$ first overcomes the nearest-neighbor magnetic coupling, and results in flips of all the spins in one layer to align with the adjacent layer, gaining energy through aligning a majority of local moments with the external field. This transition corresponds to the 1/3 fractional plateau. As the magnetic field surpasses $H_2$, the remaining unflipped moments in two separated layers flip together [Fig. 9 (c)] or flip one by one to align with the field (if the interlayer coupling is weaker than the intralayer coupling, these two ways are close in terms of energy), leading to saturation of magnetization or 2/3 kink, respectively. This mechanism naturally explains the 1/3 fractional magnetization plateau observed experimentally in the 1/3 $a^*$ AFM state, and also accounts for the weak kink in the magnetization curve reported recently [15,16]. Furthermore, during this spin-flip process, enhanced scattering of conduction electrons is anticipated, leading to violent variations in both resistivity [Fig. 10(b), Appendix A] and Hall conductance [17]. Future neutron diffraction experiments under applied magnetic fields are essential to confirm the proposed magnetic structure and test this scenario.

Given the possible close connection between the fractional magnetization plateau and the quasi-one-dimensional Fermi surface, our proposed mechanism can be experimentally tested by systematically tuning the Fermi surface—via electrostatic gating or chemical doping—taking advantage of the RKKY interaction's sensitivity to the Fermi surface topology. Intriguingly, such a tuning could stabilize alternative magnetic ground states with different periodicities, potentially leading to fractional magnetization plateaus beyond the observed 1/3 value. More broadly, this mechanism offers a tunable platform for exploring novel phenomena associated with fractional magnetization plateaus in correlated electron systems.

## CONCLUSION

In conclusion, we present a comprehensive study of the electronic and magnetic structure of the anisotropic kagome magnet $TbTi_3Bi_4$, integrating ARPES measurements with NPD, STM and DFT calculations. Our systematic temperature-dependent investigation in momentum space reveals a dual modulation of the band structure characterized by the wavevector (approximately 1/3, 0.28, 0) in the AFM state. Notably, the associated band folding leads to the formation of Dirac cones at the BZ boundary, hinting at the realization of magnetic-order-induced topological transition. By combining multiple experimental and theoretical approaches, we identify these dual band reconstructions as manifestations of the AFM order, driven by a (approximately 1/3, 0.28, 0) nesting instability. Very importantly, our DFT calculations demonstrate a direct correlation between the Tb $5d_{xz}$ orbital and the observed surprising momentum-dependent band reconstruction in the AFM state. This orbital-selective band reconstruction calls for further research into its potential role in the rich magnetic properties of $TbTi_3Bi_4$ and other materials in the $Ln$$Ti_3Bi_4$ family. Our finding provides key insights into the nature of the AFM state and the unusual fractional plateau of $TbTi_3Bi_4$, and opens avenues for manipulating magnetism through the orbital-selective control of the electronic structure.

## METHODS

### A. Single crystal synthesis and Sample characterization

$TbTi_3Bi_4$ single crystals were grown using a molten Bi flux method. High purity Tb pieces (Thermo Fisher Scientific 99.9%), Ti granules (Thermo Fisher Scientific 99.99%), and Bi granules (Aladdin, 99.999%) in a ratio of 2:3:12 were loaded in an alumina crucible and sealed in a quartz ampule. Then, the quartz ampule was heated up to 1050 °C at a rate of 200 °C/h. After annealing at this temperature for more than 30 h, the ampule was slowly cooled down to 500◦C at a rate of 2 °C/h. After removing the excess Bi flux using a centrifuge, single crystals with typical dimension 2 × 2 × 0.5 $mm^3$ were obtained, which showed a hexagonal shape.

The crystal structure of $TbTi_3Bi_4$ single crystals was checked via XRD at room temperature using an Empyrean x-ray diffractometer with Cu Kα radiation (Panalytical). Consistent with previous results, we confirm that our $TbTi_3Bi_4$ crystals crystallize in a centrosymmetric cubic structure (space group *Fmmm*, No. 69). The XRD measurements performed on a $TbTi_3Bi_4$ single crystal reveal a series of (00*l*) reflections [Fig. 1(b)], confirming the single-crystal nature of the $TbTi_3Bi_4$ sample and its *c*-axis is perpendicular to the crystal plane.

Magnetization and electrical transport measurements were performed in a Quantum Design magnetic properties measurement system and physical property measurement system, respectively. For the resistivity measurements, the electric current was applied within the ab plane, while the magnetic field was applied along the c axis.

### B. ARPES measurement

ARPES measurements were performed at BL-09U and BL-03U of the Shanghai Synchrotron Radiation Facility (SSRF), and at BL 5-2 of the Stanford Synchrotron Radiation Lightsource (SSRL), all equipped with a SCIENTA Omicron DA30L analyzer. The energy resolution at BL-09U and BL-03U (SSRF) was 10 meV - 60 meV, with an angular resolution of 0.1°. The energy and angular resolutions at BL 5-2 (SSRL) were 5 meV – 20 meV and 0.1°, respectively. Samples were cleaved *in situ* and measured under a base vacuum better than $5 \times 10^{-11}$ Torr.

### C. DFT calculation

DFT calculations are performed within the Vienna *ab-initio* simulation package [39]. The Perdew-Burke-Ernzerhof-type generalized-gradient approximation [40] is used to mimic the electronic exchange-correlation interaction. The projector augmented wave potentials with nine valence electrons for Tb (*f* electrons are not considered), four valence electrons for Ti, and five valence electrons for Bi are employed. The crystal structure of $TbTi_3Bi_4$ is fully relaxed until the remaining force on all atoms is no larger than 1 meV/Å. The energy cutoff for the plane wave basis set is 250 eV. A *k*-mesh of 6×6×6 is utilized to sample the reciprocal space. Spin-orbit coupling is considered for all calculations except for the structural relaxation. The Fermi surface is calculated by the Wannier Hamiltonian [41] extracted from DFT calculations with the basis set composed of Tb *d*, Ti *d*, and Bi *p* orbitals.

### D. NPD measurement

NPD measurements were performed using the time-of-flight (TOF) diffractometer general purpose powder diffraction (GPPD) at the Chinese Spallation Neutron Source. The GPPD provides a high-intensity neutron beam with excellent spatial resolution, making it well-suited for detailed analysis of complex magnetic structures. The sample was mounted in a vanadium holder and placed in a liquid-helium cryostat capable of reaching temperatures as low as 2 K. Diffraction patterns were collected in TOF mode over a wavelength range of 0.1–4.9 Å. The resulting NPD data were analyzed using the Rietveld refinement package within the FULLPROF suite.

## ACKNOWLEDGEMENTS

We are grateful to Chun Lin, Jianyang Ding, Jiayu Liu and Gexing Qu for their assistance with our ARPES measurements. We are also grateful to Mojun Pan for helpful discussions. B.L. acknowledges support from the Ministry of Science and Technology of China (Grant No. 2023YFA1407400), the National Natural Science Foundation of China (Grant No. 12374063), the Shanghai Natural Science Fund for Original Exploration Program (Grant No. 23ZR1479900), and the Cultivation Project of Shanghai Research Center for Quantum Sciences (Grant No. LZPY2024). H. D. acknowledges support from the New Cornerstone Science Foundation (Grant No. 23H010801236), Innovation Program for Quantum Science and Technology (Grant No. 2021ZD0302700). Y. B. H. acknowledges support by the Shanghai Committee of Science and Technology (Grant No. 23JC1403300), and the Shanghai Municipal Science and Technology Major Project. Z.W. is supported by the US Department of Energy, Basic Energy Sciences Grant No. DE-FG02-99ER45747. We gratefully acknowledge the technical support provided by the BL09U (31124.02.SSRF.BL09U) and BL03U (31124.02.SSRF.BL03U) beamlines at the Shanghai Synchrotron Radiation Facility (SSRF) for conducting the ARPES and PPMS measurements.